\documentclass[%
 reprint,
 amsmath,amssymb,
 aps,
]{revtex4-2}

\usepackage{graphicx}
\usepackage{dcolumn}
\usepackage{amsmath}
\usepackage{bm}
\usepackage{xcolor}
\usepackage{array}
\usepackage{wrapfig}

\newcommand{\Lim}[1]{\raisebox{0.5ex}{\scalebox{0.8}{$\displaystyle \lim_{#1}\;$}}}
\begin{document}

\preprint{APS/123-Quantum}
\title{
Non-uniform magnetic field as a  booster for quantum speed limit: faster quantum information processing}

\author{Srishty Aggarwal$^1$}
\email{srishtya@iisc.ac.in}
\author{Subhashish Banerjee$^2$}
\email{subhashish@iitj.ac.in}
\author{Arindam Ghosh$^1$}
\email{arindam@iisc.ac.in}
\author{Banibrata Mukhopadhyay$^1$}
\email{bm@iisc.ac.in}

\affiliation{1. Department of Physics, Indian Institute of Science, Bangalore 560012, India\\2. Department of Physics, Indian Institute of Technology Jodhpur, Jodhpur 340211, India}

\date{\today}

\begin{abstract} 
We probe the quantum speed limit (QSL) of an electron when it is trapped in a non-uniform magnetic field. We show that the QSL increases to a large value, but within the regime of causality, by choosing a proper variation in magnetic fields. We also probe the dependence of QSL on spin of electron and find that it is higher for spin-down electron in the relativistic regime. This can be useful in achieving faster speed of transmission of quantum information. Further, we use the Bremermann--Bekenstein bound to find a critical magnetic field that bridges the gap between non-relativistic and relativistic treatments and relates to the stability of matter. An analytical framework is developed. We also provide a plausible experimental design to supplement our theory.
\end{abstract}
\maketitle
\section{Introduction} 
Quantum speed limit (QSL) provides a quantitative estimate about the speed with which quantum information is processed \cite{Deffner2017}. 
Its historical roots are entrenched in the foundations of quantum mechanics. Thus, the first appearance of QSL emerged in the context of the energy-time uncertainty relation \cite{mandelstam1945energy}. 
QSL time sets the lower bound for the evolution time between two quantum states. Motivated by the Heisenberg energy-time uncertainty principle,  Mandelstam, Tamm \cite{mandelstam1945energy} and Margolus, Levitin \cite{MargolousLevitin} derived bounds on the minimum time needed for a quantum system to evolve between the states. These were combined to provide a tight bound on the QSL time for a closed quantum system. Originally developed for evolution connecting two orthogonal
states, they were subsequently generalized for arbitrary initially mixed and also between non-orthogonal states \cite{lloyd2003}. Another approach, based on the geometrical distance between the states, was recently developed \cite{Deffner2013}.


The definition of QSL in the context of open quantum systems \cite{sbbook} was developed in the last decade \cite{plenio,davidowhich,deffner}. The concept of QSL has been used to shed light into various facets of quantum information \cite{henri,paulson}, open systems \cite{xu,MarvianLidar,marvian,kosloff}, control of quantum systems \cite{campbelldeffner} and quantum thermodynamics \cite{Bekenstein,funo}. Further, using causality and thermodynamics, the important Bremermann-Bekenstein bound \cite{Bremermann,Bekenstein1990} relates the energy cost per bit of information to the QSL time.
Another fundamental issue to which the notion of QSL can be put to use is the inherent stability of the quantum state \cite{Liebbook}.

In recent times, the cross-fertilization of quantum information ideas with relativistic quantum mechanics has been particularly fruitful. Relativistic quantum simulations have impacted  developments in Leggett-Garg inequalities \cite{sbLG1,sbLG2}, probes of curved spacetime \cite{sbBani}, geometric phase \cite{sbKhushboo} and coherence \cite{sbKhushboo2} in the context of subatomic particles, such as, neutrinos and neutral mesons. It has also led to investigations into the Unruh effect \cite{sbUnruh}. Further, in a recent work \cite{eccles2021speeding}, the role of nonlocality on the rate of information spreading, as characterized by the {\it butterfly velocity},  was studied and was shown to increase with larger fields. 

Boghosian and Taylor were first to suggest the study of relativistic quantum systems with quantum simulators \cite{BOGHOSIAN199830}. Quantum simulations of Dirac particles have been proposed and investigated using a variety of systems \cite{Nori} that include single trapped ions \cite{Lamataetal}, neutral atoms \cite{Goldman} and graphene \cite{Katsnelson2006}. Graphene provides a test-bed for relativistic quantum system for quantum information  processing. A graphene based quantum capacitor has been developed that is potentially useful in producing qubits \cite{Khorasani2017}. Recently, quantum logic gate has been proposed where qubits are encoded with surface plasmons in graphene nanostructures \cite{AlonsoCalafell2019}. Further, uniform magnetic fields have been used to control the dynamics of a chain of molecular qubits \cite{Santini}. Also, magnetostrain-driven quantum engines have been proposed on graphene flakes \cite{magneticGraphene}. Thus, relativistic systems as well as magnetic field provide important tools  for quantum information processing. This sets the scene for the present investigation, where we use the relativistic dynamics of electron in presence of magnetic field to evaluate QSL.



It was shown \cite{villamizar} that in the presence of 
uniform magnetic field ($B_0$), the QSL increases and saturates to $0.2407c$, where $c$ is the speed of light, with increasing $B_0$ 
in case of spin-up electron for the superposition of ground and first excited states. Is it possible to achieve higher QSL that could help in improving the rate of quantum information processing and
making optimal control of quantum systems? What if, the magnetic field varies? There are 
many realistic scenarios, for e.g., condensed matter experiments, plasma,  astrophysical systems, where magnetic field is non-uniform for all practical purposes.
In condensed matter, examples for spatially varying magnetic field include nuclear magnetic resonance (NMR) imaging and systems with local magnetic order. In fact, NMR has been
among the first experimental methods used to implement small
quantum algorithms \cite{PengEtAl2009, Zhang2012}. Further, magnetic nanoparticles, topological defects, as well as Coulomb interaction effects can be engineered to create long-range spatially varying magnetic induction \cite{Ali,Kimberly,Tutuc}. We will touch upon one such feasible experimental design in this work.  In a star, the magnetic field is generally expected to be growing from its 
surface to center, i.e., a spatially decaying field. Even our own planet’s magnetic field is not uniform. A recent evidence shows the role of quantum physics in affecting a key reaction in a cell that enables the migratory birds to navigate using non-uniformity of Earth’s magnetic field \cite{birds}. 
We will show that QSL decreases in a spatially decaying non-uniform field. However, in a spatially growing field, it increases. 
This possibly could be beneficial in achieving a faster speed of transmission of quantum information. 

The role of critical magnetic field in determining the (non-)relativistic regimes is widely known. For a uniform magnetic field, it is given by $m_e^2c^3/\hbar e$ \textit{G} (Gauss) for electron and obtained when the gyromagnetic radius is of the order of Compton wavelength of electron, where $m_e$ and $e$ are respectively the mass and charge of electron, $c$ is the speed of light and $\hbar$ is the reduced Planck constant. However, for a non-uniform magnetic field, such an analysis is much more involved. We present here a unique way to determine the critical magnetic field for a variable magnetic field using a quantity that limits the maximal rate of information production, i.e., the Bremermann-Bekenstein bound.


In the next section, we introduce the model and the underlying framework to approach our present goal. Subsequently, we discuss the variation of QSL with respect to various parameters in \S\ref{results}. Further, in \S\ref{BBBound}, the Bremermann-Bekenstein bound is explored to bridge the gap between non-relativistic and relativistic regimes. To provide an analytical framework for our results, an ansatz is explored in \S\ref{sec:thory}. We provide the plausible experimental design to achieve variable magnetic field in \S\ref{exp} before we conclude in \S\ref{concl}. Some of the calculational details are relegated to the Appendix.

\section{Model and Framework}
For simplicity, we take a power law variation of magnetic field, given by 
\begin{equation}
\textbf{B} = B_{0}\rho^n \hat{z},
\label{eq1}
\end{equation}
in cylindrical coordinates ($\rho,\phi,z$), where `$n$' is the magnetic non-uniformity index. Note that $n>-1$ so that effective potential is always attractive \cite{SciSris}.
In this work, we take $\rho$ in $pm$ (picometer) and magnetic field $\textbf{B}$ in $G$. For dimensional consistency, $B_0$ is considered in units of $G\:pm^{-n}$. Thus, $B_0=|\textbf{B}|$ at $1\:pm$. Also, the size of the system is comparable to the gyromagnetic radius of electron.

Using a gauge freedom for the vector potential \textbf{A}, we choose
\begin{equation}
\textbf{A} = B_0\frac{\rho^{n+1}}{n+2}\hat{\phi}=A\hat{\phi}.
\label{eqn2}
\end{equation}

To obtain eigenstates of an electron and their corresponding eigen-energies, we solve the Dirac equation 
\begin{equation}
i\hbar\frac{\partial\boldsymbol{\Psi}}{\partial t} = \left[ c\boldsymbol{\alpha}\cdot\left(-i\hbar\textbf{$\nabla$}-\frac{q\textbf{A}}{c}\right) + \beta m_{e}c^2\right]\boldsymbol{\Psi},
\label{eqn3}
\end{equation}
where \textit{$m_e$} and \textit{q} are the mass and charge of electron respectively, \textbf{$\alpha$} and $\beta$ are Dirac matrices and \textbf{A} is the vector potential.

We begin with relativistic calculations, i.e., the solution of Dirac equation, so that they reduce to non-relativistic scenario while considering weak magnetic fields. This will enable us to cover a wide range of magnetic field strengths. The solution of Dirac equation for the power-law magnetic field given by Eq. (\ref{eq1}) has been explored in detail earlier \cite{SciSris}, part of which will be employed for the present purpose.

Let
\begin{equation}
\boldsymbol{\Psi}(t,r)= e^{\frac{iEt}{\hbar}}\psi(r),
\end{equation}
then the general solution of $\psi$ in presence of such a  magnetic field is given by
\begin{equation}
\psi = e^{i\left(m\phi+\frac{p_{z}}{\hbar}z\right)}\begin{bmatrix}
R(\rho)\\
-R(\rho)\\
\end{bmatrix},
\label{eqn4}
\end{equation}
where $R(\rho)$ is the two-component matrix, \textit{`m$\hbar$'} is the angular momentum of the electron and $p_z$ is the eigenvalue of momentum in the z-direction. The method of obtaining $R(\rho)$ numerically is discussed in detail in the Appendix A. We have chosen $p_z$ and $m$ both equal to zero in this work.

\begin{figure*}
	\includegraphics[scale=.5]{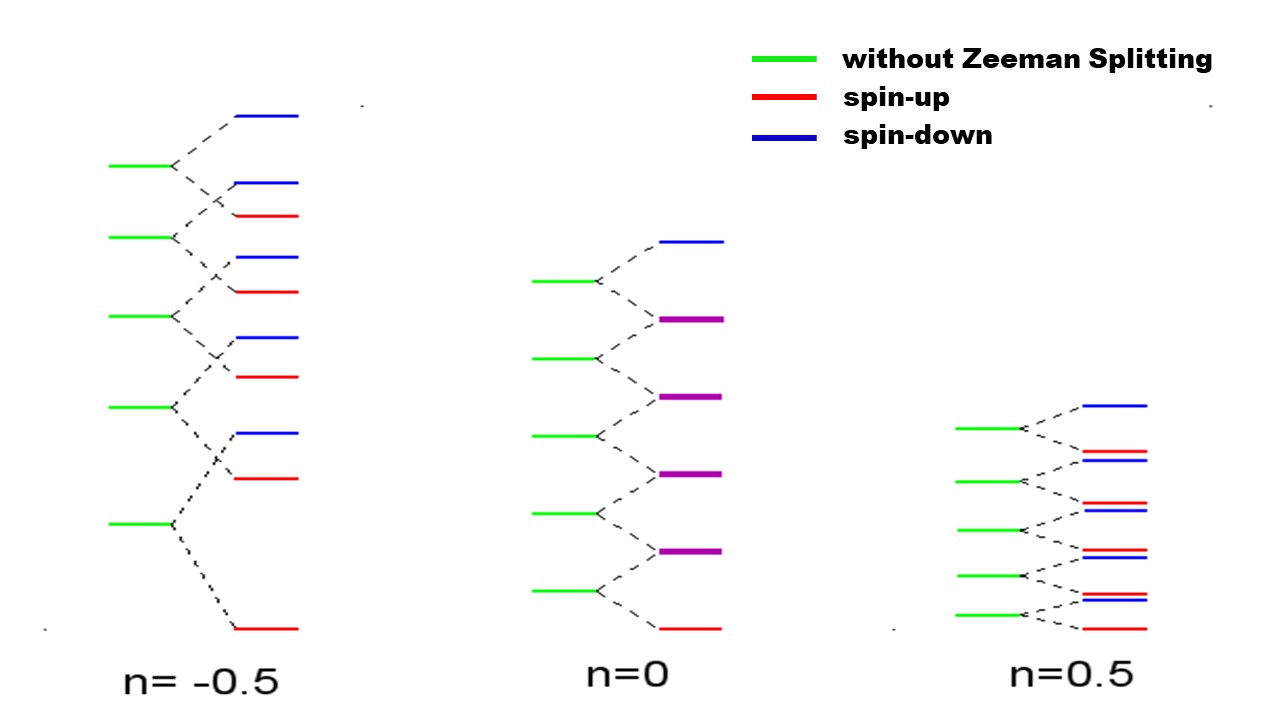}
\caption{Comparison of eigenlevels and the split states of $+\sigma.B\:{\rm and}\:-\sigma.B$ for $n=-0.5,0$ and $0.5$. }
\label{fig1}
\end{figure*}

The energy of a level $\nu$ is given by
\begin{equation}
E_{\nu} = m_{e}c^2\:\sqrt{1+\alpha_{\nu}},
\label{Eq:Ealpha}
\end{equation}
where $\alpha_{\nu}$ represents the eigenvalue for the level $\nu$ (see Appendix A).

The eigen--spectrum, corresponding to $\alpha_{\nu}$, for five levels is given in FIG. \ref{fig1}. It is interesting that with increasing eigen states, eigenvalue difference between the two consecutive states remains same for $n=0$, decreases for $n<0$ and increases for $n>0$. Further, the spin degeneracy that is present in the case of uniform magnetic field is lifted in the presence of non-uniform magnetic field. The alignment of spin-up and spin-down levels is also different for positive and negative $n$. For positive $n$, the lower level of spin-up electron is always below the higher level of spin-down electron, whereas, for negative $n$, the former lies above the latter.

To evaluate the QSL of an electron for the evolution from one state to the other, we require the information about two main quantities:
\begin{enumerate}
\item \emph{Radial displacement of electron ($\rho_{disp}$)}, 
\item \emph{Minimum time of evolution ($\tau_{QSL}$)}.
\end{enumerate}

Let us assume that the particle is in the superposition of the $\nu_{th}$ and $(\nu +1)_{th}$ states at all times, such that its initial ($t=0$) and final ($t=\tau_{QSL}$) states are
\begin{equation}
\Psi(r,0) = \frac{1}{\sqrt{2}}\left[\psi_{\nu}(r)+\psi_{\nu+1}(r)\right],
\end{equation}
and
\begin{equation}
\Psi(r,\tau_{QSL}) = \frac{1}{\sqrt{2}}\left[\psi_{\nu}(r)e^{\frac{iE_{\nu}\tau_{QSL}}{\hbar}}+\psi_{\nu+1}(r)e^{\frac{iE_{\nu+1}\tau_{QSL}}{\hbar}}\right].
\end{equation}

The mean radial position at a time $t$ is given by
\begin{equation}
\langle\rho\rangle = \frac{1}{2}\left[\langle\nu|\rho|\nu\rangle + \langle\nu+1|\rho|\nu+1\rangle + 2\langle\nu|\rho|\nu+1\rangle cos({\cal{E}}t)\right],
\end{equation}
where 
\begin{equation}
\nonumber
{\cal{E}} = \frac{E_{\nu +1}-E_{\nu}}{\hbar}.
\end{equation}

Hence, the radial displacement of electron is 
\begin{align}
\rho_{disp} &= |\langle\rho\rangle_{\tau_{QSL}}-\langle\rho\rangle_0|\nonumber \\ 
&=2\left\vert\int^\infty_0 \rho D_S(\rho)d\rho\: \right\vert,
\label{eqn5}
\end{align}
where 
\begin{equation}
D_s(\rho)=\psi^{\dagger}_{\nu}\:\rho\:\psi_{\nu+1}
\label{eqn6}
\end{equation}
($\rho$ in Eq. (\ref{eqn6}) is due to cylindrical volume element $\rho d\rho d\phi dz $). 

Since, for this system, the initial and final states of the electron are orthogonal, Mandelstam-Tamm (MT) \cite{mandelstam1945energy} and Margolous-Levitian (ML) \cite{MargolousLevitin} bounds become same. The minimum time of evolution, given by MT bound, is
\begin{equation}
\tau_{QSL}=\frac{\pi \hbar}{2\Delta H},
\end{equation}
with
\begin{equation}
\Delta H =\frac{ E_{\nu +1}-E_{\nu}}{2}.
\end{equation}
Thus, QSL of an electron is given by
\begin{equation}
\tilde{v} = \frac{\rho_{disp}}{\tau_{QSL}}.
\label{eq vel}
\end{equation}

\section{\label{results}Results} 
The QSL of spin-up electron was probed in constant magnetic field earlier \cite{villamizar}. We investigate the variation of QSL of electron with different parameters in presence of variable magnetic field for both spin-up as well as spin-down electrons. 
\subsection{\label{spin var}Variation with spin}

\begin{figure}
\includegraphics[scale=.55]{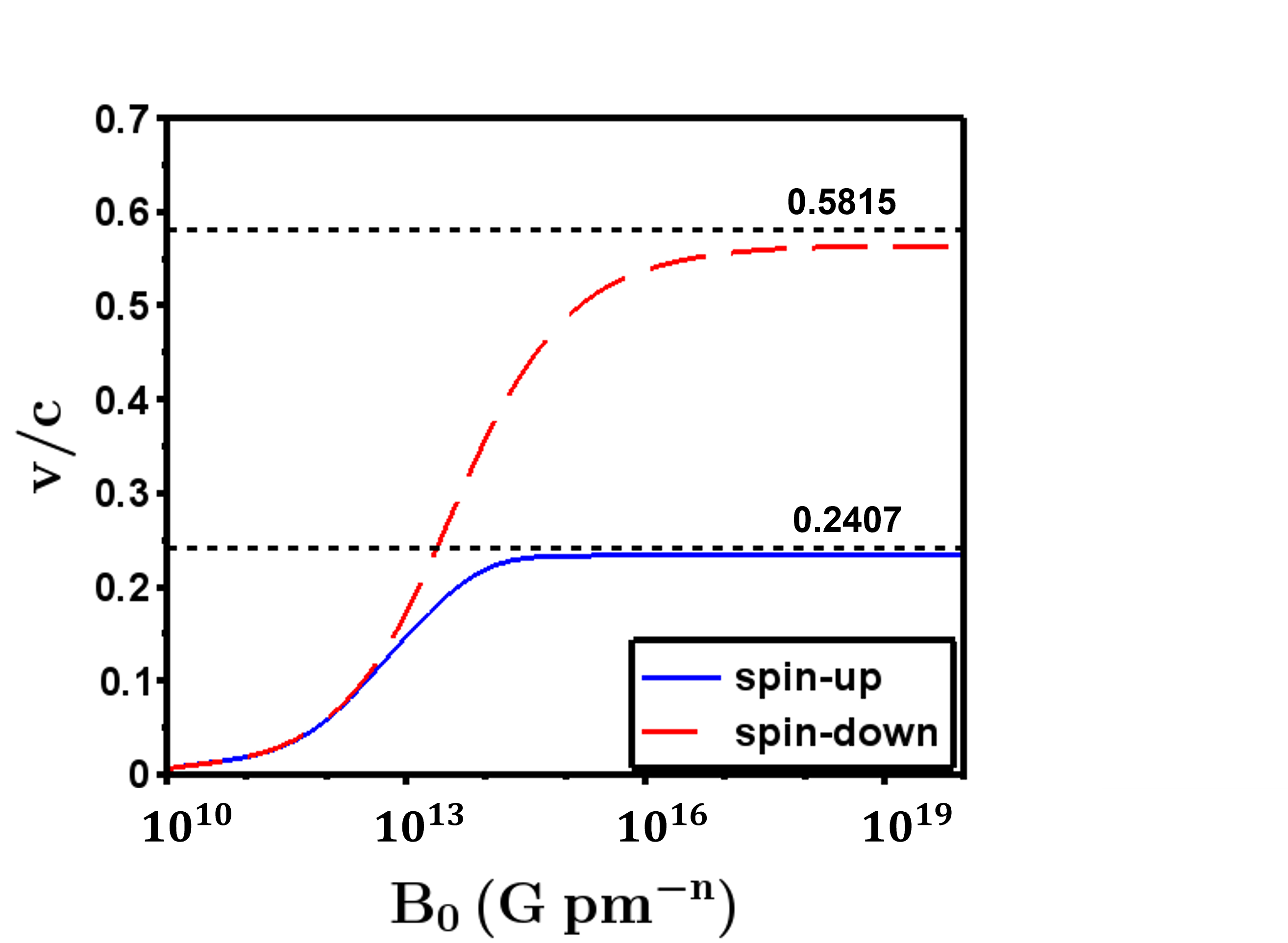}
\caption{Comparison of radial speed of spin-up and spin-down electrons for uniform magnetic field.}
\label{spinvar}
\end{figure}


For a given $n$, QSL increases with increasing magnetic field strength and then saturates. The QSL in the limit of $B_0\rightarrow\infty$ is called as saturated quantum speed limit (SQSL).
Figure \ref{spinvar} shows that the SQSL of spin-down electron is more than twice of` that for spin-up electron in presence of uniform magnetic field for the superposition of ground and first excited states.

We can calculate SQSL of a spin-down electron for a superposition between ground and first excited states for uniform magnetic fields. In the limit of $B_0\rightarrow\infty$, the energies of spin-down electron can be approximated as
\begin{equation}
E_0 = m_ec^2;\:\;  \:E_1 = m_ec^2\sqrt{\alpha_1}=m_ec^2\sqrt{\frac{2eB_0\hbar}{m_e^2c^3}}
\end{equation}
which renders
\begin{equation}
  \tau_{QSL}\approx\frac{\pi \hbar}{m_ec^2\left(\sqrt{\frac{2eB_0\hbar}{m_e^2c^3}}-1\right)}\approx\frac{\pi}{2c\beta},
  \end{equation}
  where $\beta = \sqrt{\frac{eB_0}{2\hbar c}}$.

The radial displacement is same for both the spins of electrons. Hence, radial displacement for spin-down electron (as given for spin-up electron \cite{villamizar}) can be taken as 
\begin{equation}
 \rho_{disp} = \frac{\sqrt{\pi}}{4\beta}\left(1+\frac{3}{2\sqrt{2}}\right).
\label{eq:rho_uniform}
\end{equation} 
 
Consequently, SQSL for spin-down electron $v_{lim_{\downarrow}}$ in presence of constant magnetic field is given by
\begin{equation}
v_{lim_{\downarrow}} = \frac{c}{2\sqrt{\pi}}\left(1+\frac{3}{2\sqrt{2}}\right) = 0.5815\:c.
\end{equation}
Note that the SQSL for spin-up electron $v_{lim_{\uparrow}}$ is $0.2407\:c$ \cite{villamizar}. Hence, one can achieve higher SQSL for spin-down electron.

\subsection{\label{n var} Variation with $n$}
\begin{figure*}
\includegraphics[scale=.45]{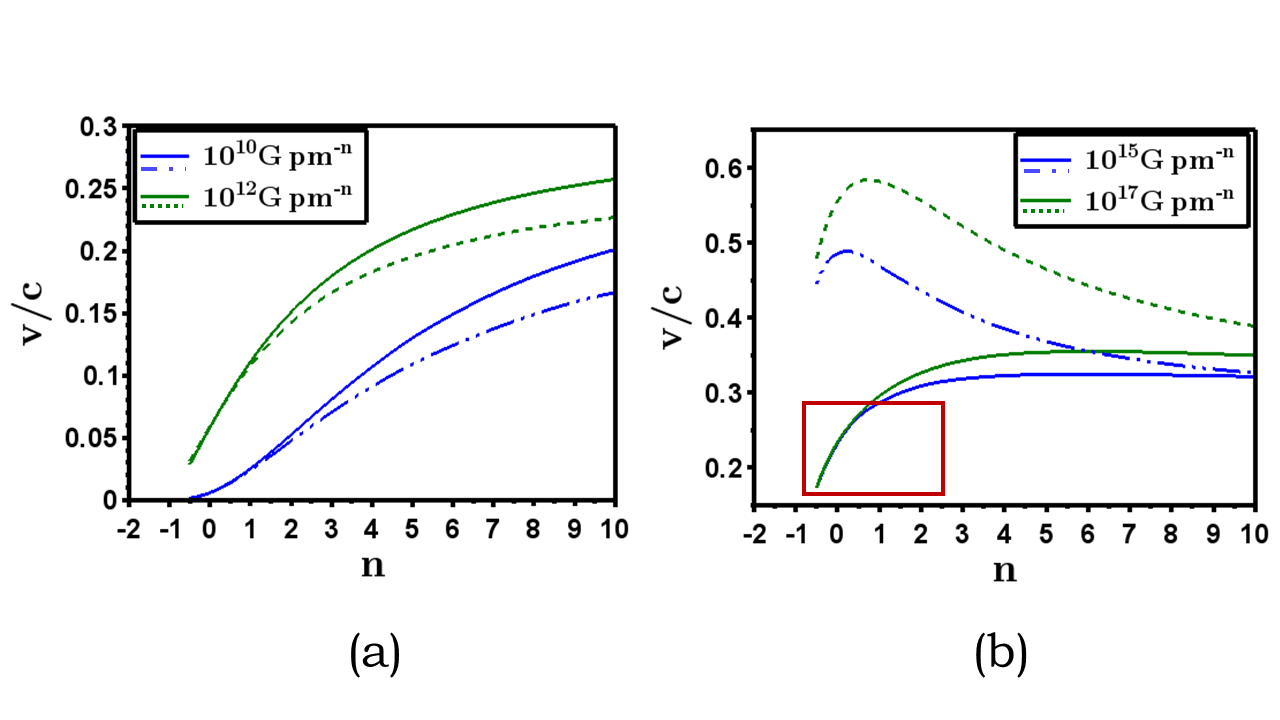}
\caption{Variation of QSL of spin-up (solid lines) and spin-down (broken lines) electrons with $n$ for different strengths of the central magnetic field, for (a) lower fields (b) higher fields. The rectangle in (b) highlights that the QSL for spin-up electron merges for low $n$ for $B_0 = 10^{15}$ and $10^{17}\:G\:pm^{-n}$, signifying that SQSL has been attained for these $n$.}
\label{nvar}
\end{figure*}

We now analyse the variation of QSL with $n$ in non-relativistic and relativistic regimes. FIG. \ref{nvar} (a) shows the variation for lower magnetic fields, wherein $B_0=10^{10}\:G\:pm^{-n}$ and $10^{12}\:G\:pm^{-n}$ can be considered as the non-relativistic regime while  $B_0>10^{14}\:G\:pm^{-n}$ as the relativistic regime, shown in FIG. \ref{nvar} (b). The variation of QSL between spin-up and spin-down electrons is quite opposite between non-relativistic and relativistic regimes. 
In the non-relativistic regime, QSL is same for lower $n$ ($\leq1$) for both the spins and becomes smaller for spin-down electron as $n$ increases. In the relativistic regime, it becomes higher at low $n$ for the spin-down case which decreases with increasing $n$. 


Further, FIG. \ref{nvar} (b) shows that as magnetic field increases, QSL tends to saturate. Note that the curves representing $10^{15}$ and $10^{17}~G~pm^{-n}$ for spin-up electron overlap for $n\lesssim0$ depicting that QSL does not increase further with increase in $B_0$, and has attained SQSL. The figure also indicates the significant difference between QSL for spin-up and spin-down electrons, thus, showing the importance of spin at small non-linearity (low $n$) for relativistic electron.
  
The variation of QSL with $n$ is steeper for lower $B_0$. Note that QSL for $n=1$ becomes four times QSL for $n=0$ at $B_0 = 10^{10}\:G\:pm^{-n}$, while at $B_0=10^{15}\:G\:pm^{-n}$, the ratio of the two is just $1.2$ for spin-up electron. 
Thus, even a linear variation of the magnetic field in laboratories, could help to attain much higher QSL of electron as compared to its constant counterpart. 

QSL for spin-up and spin-down electrons tends to be same in the non-relativistic regime 
for small variation of magnetic field ($n\leq 1$), because
\begin{align}
\Delta H(relativisic)  = m_e c^2\left(\sqrt{1+\alpha_{\nu+1}} - \sqrt{1+\alpha_{\nu}}\right) \label{sqrtalpha}\\
  \underrightarrow{non-relativistic}\:\; \frac{m_e c^2}{2}\left(\alpha_{\nu +1}-\alpha_{\nu}\right). \label{deltaH}
\end{align} 
The above set of equations shows that $\Delta H$ depends on the difference of $\alpha$ between two consecutive levels in the non-relativistic regime, which is not for the relativistic regime. Since,
the difference between the respective levels for spin-up and spin-down cases is nearly same for small variation of magnetic fields, QSL is same for both the spin orientations of electron at low $B_0$, shown in FIG. \ref{nvar}(a).


It is interesting that the variation of SQSL with $n$ is similar to the QSL variation for spin-down electron, as shown for the latter in Fig. \ref{nvar}(b), i.e., SQSL increases with increasing $n$ and attains a peak.
While for the spin-down case, SQSL is maximum for the quadratic variation of magnetic field, $n$ has to be $15$ to attain maximum SQSL for spin-up electron.

\subsection{\label{B var}Variation with $B_0$}


\begin{figure}
\includegraphics[scale=.5]{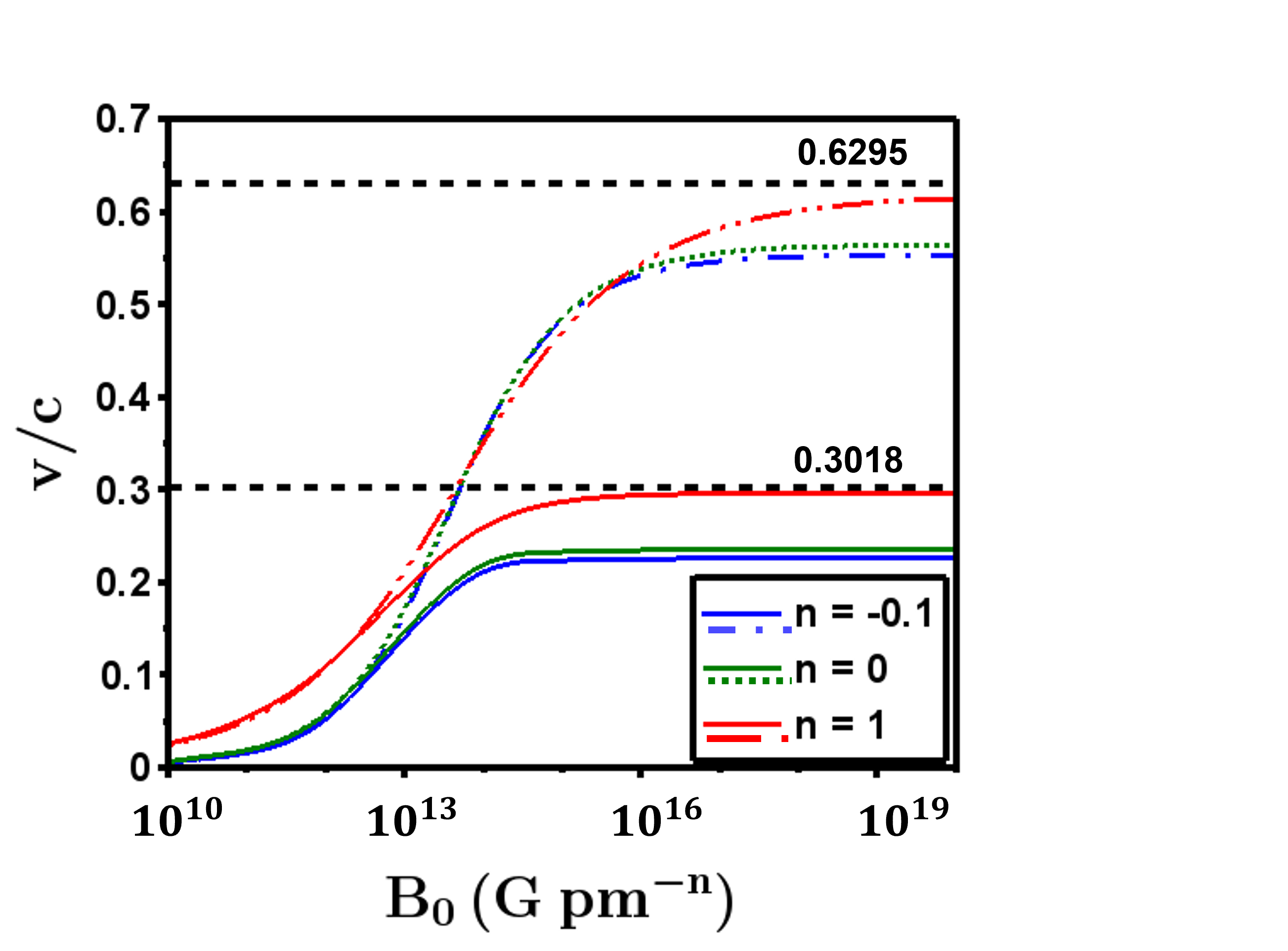}
\caption{Comparison of QSL of electron for increasing ($n=1$), decreasing ($n= -0.1$) and uniform ($n=0$) magnetic fields with different magnetic field strength ($B_0$) for spin-up (solid lines) and spin-down (broken lines) electron for a superposition of ground and first excited states. Here, SQSLs for $n=1$ are highlighted for both the spins. For $n=-0.1$, SQSLs for spin-up and spin-down electrons are $0.2288c$ and $0.5638c$, respectively, and for $n=0$, they  are $0.2407c$ and $0.5815c$ for spin-up and spin-down electrons, respectively.}
\label{Bvar}
\end{figure}
FIG. \ref{Bvar} shows that QSL increases with the increase in $B_0$ and then reaches SQSL, which is a function of $n$, for a superposition of ground and first excited states. For spin-up electron in the figure, for $n=-0.1$ and $n=0$, QSL reaches SQSL, while for $n=1$, QSL is still increasing and has not reached SQSL in the magnetic field regime shown in the figure. This indicates that 
$B_0$ leading to SQSL is larger for $n=1$ as compared to $n=0$ and $n=-0.1$. 
Similar features hold for spin-down electron.
This implies that $B_0$ corresponding to SQSL
increases with the increase in $n$.
Also, note that SQSL for spin-down electron ($v_{lim_{\downarrow}}$) is larger than that for spin-up electron ($v_{lim_{\uparrow}}$), as discussed above.

\subsection{\label{state var}Variation with state}
\begin{figure}
\includegraphics[scale=.48]{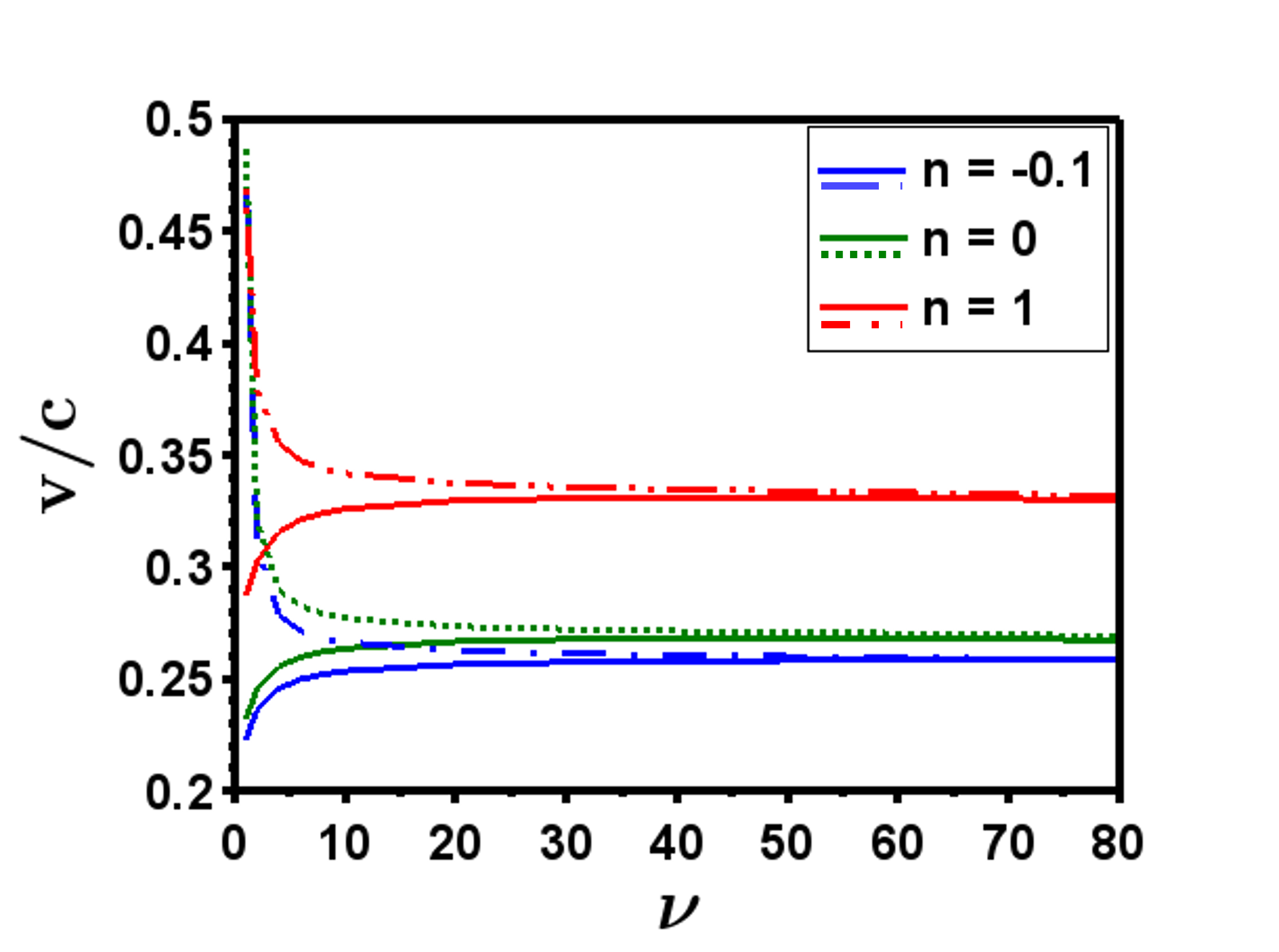}
\caption{Variation of quantum speed with different superposition of states ($\psi_{\nu},\psi_{\nu +1}$) at $B_0 =10^{15}\:G\:pm^{-n}$ for spin-up (solid lines) and spin-down (broken lines) electron.}
\label{fig state var}
\end{figure}
QSL varies with state $\nu$ in an opposite manner for opposite electron spins, considering the system as superposition of states ($\psi_{\nu}, \psi_{\nu +1}$).
For spin-up electron, it increases as energy state increases and then saturates, whereas for spin-down electron, it decreases before saturation as shown in FIG. \ref{fig state var}. However, the saturation value is independent of spin and increases with $n$.

The opposite variation for spin-up and spin-down electrons, in lower energy states owes to the large energy difference between the ground and first excited states for spin-down electron, thereby making its $\tau_{QSL}$ smaller than spin-up electron. The radial displacement is similar for both the spins. It initially increases for increasing $\nu$ and then saturates. Therefore, QSL, being strongly dependant on $\tau_{QSL}$, is significantly higher for spin-down electron for superposition of ground and first  excited states. However, for higher excited states, the energy difference between the consecutive levels becomes similar between states for both the spins, leading to the convergence of QSL with states.

\section{\label{BBBound} Bremermann-Bekenstein Bound}
Bremermann \cite{Bremermann} argued that speed, memory, and processing capacity
of any computational device are limited by certain physical
barriers: the light barrier, the quantum barrier, and the thermodynamical barrier. In an almost heuristic way, he found an upper bound on the rate of information communication by 
using Shannon’s work \cite{Shannon} on classical channel capacities and the associated noise energy, speed of light, and the energy-time uncertainty principle. Subsequently, Bekenstein proposed an alternative approach \cite{Bekenstein, Bekenstein1990} based on the upper bound on the information retrieval rate from black holes. In particular, the Bremermann–Bekenstein bound \cite{deffner} is given by
\begin{equation}
\frac{\langle H \rangle}{I} > \frac{\hbar \ln 2}{ \pi \tau_{QSL}}
\label{BB bound}.   
\end{equation}
It relates the energy cost $\langle H \rangle$ per bit of information $I$ to the QSL time. To give a meaning to the information $I$, let us consider Bekenstein's arguments \cite{Bekenstein1981}, where information $I$ is related to entropy $S$ whose upper bound in a given region of space of linear dimension $R$ can be expressed as the inequality $\frac{S}{\langle H \rangle} < \frac{2\pi k_B R}{\hbar c}$. 

\begin{figure*}
    \centering
    \includegraphics[scale=.5]{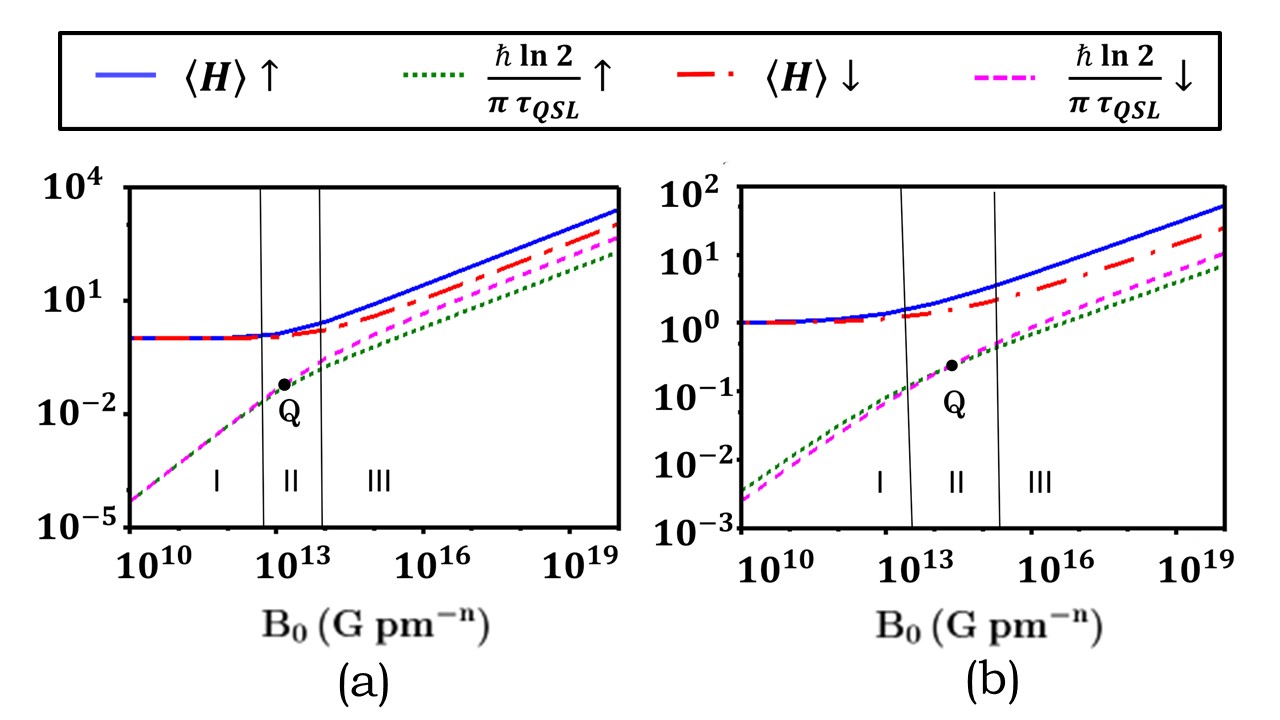}
    \caption{Transition from non-relativistic to relativistic regimes and critical magnetic field for spin-up and spin-down electrons for the variation of (a) uniform magnetic field ($n=0$), (b) non-uniform magnetic field with $n=2$.}
    \label{fig:Bstability }
\end{figure*}

The Bremermann-Bekenstein bound has been an important result in quantum information, cosmology and black hole physics \cite{Deffner2010,DeffnerPRR2020, Bekenstein, Bekenstein1990}.
Refinements to this bound, resulting in the so called generalized Bremermann–Bekenstein bound, have been made in recent times,  wherein the information $I$ is replaced by the accessible information characterized by the Holevo information \cite{DeffnerPRR2020}. This bound is also dependent on $\tau_{QSL}$. Here, we will explore how Bremermann-Bekenstein bound can be used to determine important properties of the system.

\paragraph{Determination of critical magnetic field:}
We depict the left handside (LHS) of Eq. (\ref{BB bound}) for one bit of information and its right handside (RHS) for (a) uniform magnetic field ($n=0$) and (b) non-uniform magnetic field ($n=2$) in FIG. \ref{fig:Bstability }. The bridge between the magnetic fields in the non-relativistic and relativistic regimes is clearly visible in the figure. The sub-figures are divided into three regions. In region I, $\langle H\rangle$ and $(\hbar ln2)/(\pi\tau_{QSL})$ are well separated and exhibit a tendency of convergence towards region II in weak magnetic fields, indicating \textit{non-relativistic} regime; in region II the separation reduces, depicting transition and the region III shows the two lines becoming almost parallel and is the \textit{relativistic} regime. It is known that the critical magnetic field $B_c$ above which relativistic Landau quantization is important, in realm of uniform magnetic field, is $m_e^2c^3/\hbar e=4.414\times 10^{13}G$. Note that $B_c$ lies in region II in FIGs. \ref{fig:Bstability } (a) and (b), thus, justifying our division. For uniform magnetic field considered in FIG. \ref{fig:Bstability } (a), $(\hbar ln2)/(\pi\tau_{QSL})$ is spin independent in the non-relativistic regime and is larger for spin-down electron in the relativistic regime. On the other hand, it is spin dependent even in the non-relativistic regime in presence of non-uniform magnetic field as shown in FIG. \ref{fig:Bstability } (b). It is lower for the spin-down electron in the non-relativistic regime, while higher in the relativistic one. Point \textbf{Q} ($B_0 = 1.35\times10^{14}\:G\:pm^{-n}$) in region II in FIG. \ref{fig:Bstability } (b) is the intersection point for spin-down and spin-up electrons and may represent the critical magnetic field for $n=2$. Thus, Bremermann-Bekenstein bound provides a simple method to estimate the critical magnetic field.

\paragraph{Stability of system:}
Stability of a system is generally of two types, \textit{stability of first kind} and \textit{stability of second kind} \cite{Liebbook}. The former quantifies the finiteness of the expectation value of energy for a quantum system, while the latter involves providing an understanding of how extensive quantities such as energy and volume scale with the number of atoms and requires the potential explicitly. As it turns out, the Bremermann–Bekenstein bound \cite{deffner} is bounded from below by a quantity involving $\tau_{QSL}$ (RHS). Hence, finiteness of the RHS of the Eq. (\ref{BB bound}) would imply finiteness of average energy per bit of information. This suggests the connection of $\tau_{QSL}$ to the stability of the first kind via the Bremermann–Bekenstein bound. In FIG. \ref{fig:Bstability }, we show the variation of the lower bound of Eq. (\ref{BB bound}). Its finiteness is consistent with the Bremermann–Bekenstein bound, and thereby the stability of the first kind. We hope to undertake it in greater detail in a future work.

Thus, the connection between Bremermann-Bekenstein bound and $\tau_{QSL}$ has the following implications. On one hand, based on it, one is able to uncover the spin dependence of $\tau_{QSL}$.
On the other hand, it helps in understanding the transition from the non-relativistic to the relativistic regimes. Moreover, it suggests a connection to the stability of first kind.


\section{\label{sec:thory}Analytical Ansatz}
\begin{figure}
    \centering
    \includegraphics[scale=0.4]{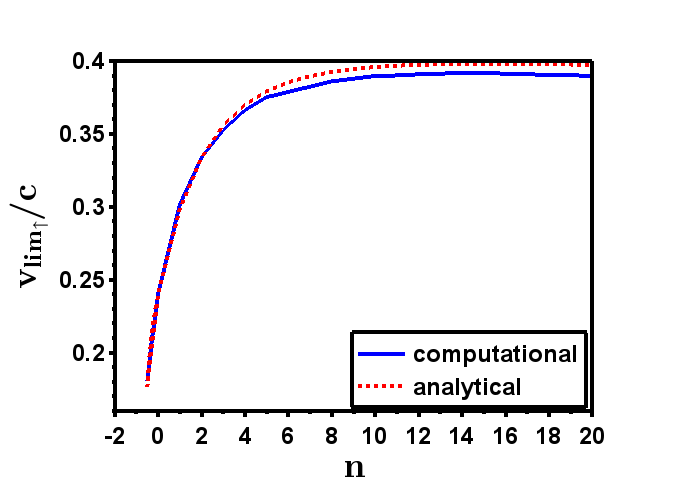}
    \caption{Comparison between the analytical and the computational $v_{lim_{\uparrow}}$.}
    \label{fig:vliml}
\end{figure}
Here, we intend to obtain the SQSL of spin-up electron for a superposition of ground and first excited states for general $n$, analytically.
Since, we do not have an analytical form of the ground and the first excited states wave functions, it is not possible to calculate the radial displacement using Eq. (\ref{eqn5}) analytically. What we know are the analytical expressions for eigenvalues \cite{SciSris}, given  by 
\begin{equation}
\alpha_\nu = C_3\: B_0^{\frac{2}{n+2}}\:(\nu+C_5)^{\frac{2+2n}{n+2}}\left[1\pm \frac{C_5}{ (\nu+C_5)}\right],
\label{alpha}
\end{equation}
such that $C_3$ is a constant which depends on $n$ and $C_5 \sim 0.5$.

We try to obtain the radial displacement of spin-up electron in presence of uniform magnetic field, given by Eq. (\ref{eq:rho_uniform}) in the limiting case of $B_0\rightarrow\infty$ \cite{villamizar}, in terms of the ground and first excited states energies and then generalise the expression for all $n$. Hence, we choose the following ansatz for the radial displacement of spin-up electron in presence of a magnetic field varying with $n$:
\begin{align}
    \rho_{disp} &= {\left[\Gamma\left(\frac{2}{n+2}\right)\right]}^{-\frac{1}{n+2}} \Gamma \left(\frac{3}{n+2}\right) \lambda_e \frac{(E_{1}+E_0)^2}{E_{1}^3}\\
    &={\left[\Gamma\left(\frac{2}{n+2}\right) \right]}^{-\frac{1}{n+2}} \Gamma \left(\frac{3}{n+2}\right) \lambda_e \times\nonumber\\
    &\qquad\qquad\qquad\frac{(\sqrt{1+\alpha_1}+\sqrt{1+\alpha_0})^2}{(\sqrt{1+\alpha_1})^3},
    \label{eq:rho_alpha}
\end{align}
where $\lambda_e = \hbar / m_e c$ and $\Gamma (x)$ is the standard gamma-function given by
\begin{equation}
    \Gamma \left(\frac{a}{b}\right) = b \int_0^{\infty} x^{a-1}e^{-x^b}dx.
\end{equation}


Therefore, to evaluate SQSL of spin-up electron for the superposition of ground and first excited states in such a magnetic field, a very high value of $B_0$ can be considered such that $\alpha_1$, $\alpha_0 >> 1$. Then
\begin{equation}
     \Lim{B_0 \rightarrow\infty} \rho_{disp} \sim \left[\Gamma\left(\frac{2}{n+2}\right)\right]^{-\frac{1}{n+2}} \Gamma \left(\frac{3}{n+2}\right) \lambda_e \frac{(\sqrt{\alpha_1}+\sqrt{\alpha_0})^2}{(\sqrt{\alpha_1})^3}.
\label{rho_lim}
\end{equation}

In this limit, for uniform magnetic field, $E_0$ and $E_1$ tend to $2m_ec^2\lambda_e \beta$ and $2\sqrt{2}m_ec^2\lambda_e \beta$ respectively from Eq. (\ref{lim energy uniform}). Substituting these values and $n=0$ in the above equation, we get back Eq. (\ref{eq:rho_uniform}), thus, confirming the validity of the above ansatz.

Further, 

\begin{equation}
    \Lim{B_0\rightarrow\infty}\tau_{QSL} \sim \frac{\pi \lambda_e}{(\sqrt{\alpha_1}-\sqrt{\alpha_0})}.
\label{t_lim}
\end{equation}
Substituting the values of $\alpha_1$ and $\alpha_0$ from Eq. (\ref{alpha}) in Eqs. (\ref{rho_lim}) and (\ref{t_lim}) and taking their ratio, SQSL ($v_{lim_{\uparrow}}$) is given by
\begin{multline}
     v_{lim_{\uparrow}} = {\left[\Gamma\left(\frac{2}{n+2}\right)\right] }^{-\frac{1}{n+2}} \Gamma \left(\frac{3}{n+2}\right)\times\\
     \frac{(F(1)+F(0))^2(F(1)-F(0))}{\pi\:F(1)^3},
\label{eq:vqsl}
\end{multline}
where
\begin{equation}
    F(\nu) = \sqrt{(\nu+0.5)^\frac{2+2n}{2+n}+0.5\:(\nu+0.5)^\frac{n}{n+2}}.
\end{equation}

Fig. \ref{fig:vliml} shows the comparison between the analytical SQSL and computational QSL evaluated at very high $B_0$ (which we choose hypothetically $~10^{30}\:G\:pm^{-n}$). Here, analytical and computational values merge at low $n$. As $n$ increases, $\alpha_0$ and $\alpha_1$ approach $\sim1$ at this magnetic field due to the dependence of $\alpha_{\nu}$ on $B_{0}^{\frac{2}{n+2}}$, as shown in Eq. (\ref{alpha}). Thus, an even stronger magnetic field is required to attain SQSL for large $n$. Hence, Eq. (\ref{eq:vqsl}) provides an upper limit of QSL for spin-up electron, for all $n$. Also, SQSL never crosses $0.4c$, thereby, providing the upper limit of QSL for the superposition of ground and first excited states, in any kind of magnetic field independent of $n$.   

\section{\label{exp}Experimental Design}
\begin{figure*}
    \centering
    \includegraphics[scale=0.5]{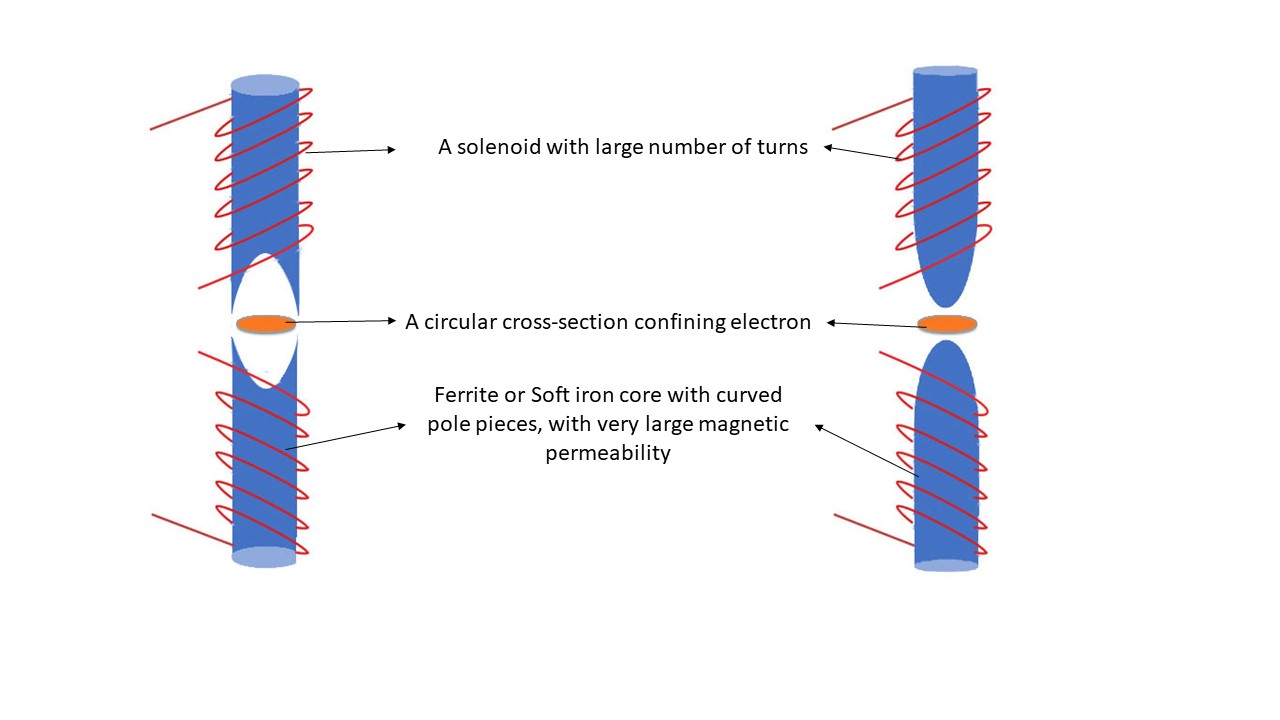}
    \caption{Model for attaining the variable magnetic field. The left one with concave pole pieces can be used to attain increasing magnetic fields ($n>0$) and the one on the right having convex poles is suitable for decreasing magnetic fields ($n<0$).}
    \label{fig:experiment}
\end{figure*}

The  proposed  power  law  variation  in  magnetic  field can be achieved in a laboratory environment in multiple ways, depending on the nature and spatial scale of the quantum system under investigation. For instance, we can take a solenoid with its core having curved pole pieces as shown in FIG. \ref{fig:experiment}. Ferrite or soft iron can be used as core, since they have high magnetic permeability which is desirable to achieve strong magnetic fields. The overall shape of the pole pieces is cylindrical, but the concavity (convexity) at the end of the pole pieces will make magnetic field to decrease (increase) strongly at the center, but it will be large (small) at the boundary. The nature of concavity (convexity) can be engineered to achieve various values of $n$.

As an illustration, if we take a solenoid with $100$ turns per cm, ferrite core with concave pole pieces of diameter 1 mm and carrying 1 A current, we can generate magnetic field close to $10^4\:G$ at the edge with a spatially increasing magnetic field. The sample calculation, of attaining non-uniform magnetic field using this design, is given in Appendix B. Let us assume that the electron is confined to the two dimensional circular plane within $1\:\mu m$ between the pole pieces. 
If the magnetic field has linear variation from zero at the centre to a maximum value of $10^4\:G$ at the edge of ferrite core, it corresponds to $10\:G$ at the boundary of the circular plane of electron. Hence, $|\textbf{B}|=10~G$ at $\rho=0.5\:\mu m=5\times10^5\:pm$ when $n=1$. Thus, using Eq. (\ref{eq1}), $B_0 = 10/(5\times10^{5})=2\times10^{-5}~G~pm^{-n}$. Correspondingly, QSL for spin-up and spin-down electrons is $~3.2\times 10^{-7}c$ and $~3\times 10^{-7}c$, respectively. A consideration of the same situation for uniform magnetic field with $B_0=10\:G$ yields a QSL of $1.9\times 10^{-7}c$ for both the spin orientations of electron at the boundary of the circular plane mentioned above. The size of this plane generally corresponds to the scale of superconducting Josephson junction \cite{2019Natur.569...93R} or a lithographically defined quantum dots on two dimensional electron system in semiconductors \cite{Kouwenhoven_2001}. 
Spatially varying magnetic field can also be achieved in a three dimensional architecture, where the quantum system is placed between a pair of lithographically patterned films of strong perpendicular magnetic anisotropy \cite{doi:10.1063/1.4749818}. The spatial variation of the magnetic field in the latter case can be realized in multiple ways, for example, through radial variation in film thickness. Therefore, increased QSL in presence of spatially increasing magnetic fields can be attained experimentally. We leave the experimental verification of these results for a future work. 

\section{\label{concl}Conclusions}

Attaining higher quantum speed would be a desirable attribute for quantum information processing. We suggest, in this work, variable magnetic fields as a possible solution over constant magnetic fields to achieve higher QSL of electron, in general, charged fermions. 

We have computed the QSL in presence of variable magnetic field for spin-up and spin-down electrons. For the present purpose, we have chosen power law variation of the magnetic field, where its magnitude varies in the plane, to which the electron is confined, but has a constant direction perpendicular to the plane. We have shown that electron can attain high QSL in presence of spatially increasing magnetic field for both the spins of electron within the regime of causality. However, only in the relativistic regime, a spin-down electron can attain almost twice the value of QSL than spin-up electron, whereas, in the non-relativistic regime, QSL for latter is higher or equal to the former. 
We have found an analytical expression for the estimation of the maximum attainable speed of a spin-up electron for the superposition of ground and first excited states. Spin-up electron cannot attain speed greater than 0.4c in presence of magnetic field irrespective of its variation, for our chosen field profile. We have further explored the usefulness of the Bremermann-Bekenstein bound in determining the critical magnetic field, above which non-relativistic treatment fails, and its plausible connection to the stability of matter. 

From the \textit{foundational} point of view, this work not only highlights the role of relativistic treatment in the field of quantum information but also allows us to probe various facets of relativistic dynamics of electron in a generally non-uniform magnetic field using QSL. On one hand, the relativistic treatment allows for causality bounds for both the spin-up and spin-down electrons. On the other hand, it results in the concept of a critical magnetic field which enables looking at non-relativistic and relativistic treatments from a uniform perspective.

We have provided a plausible \textit{experimental} design for a laboratory implementation of the ideas. This centers around creating (rapidly) spatially varying magnetic fields. Nevertheless,
there exist multiple ways of obtaining the same. Local magnetism can be engineered with controlled spatial dependence along grain boundaries or topological defects in graphene-like two-dimensional materials, for example, the transition metal dichalcogenides \cite{ACS}. We have compared the QSL for uniform and non-uniform magnetic fields in the framework of proposed experimental set-up which is consistent with the theory developed here. 

The current study possesses a high \textit{outreach} potential. It has the scope for applications in burgeoning fields such as, relativistic quantum thermodynamics \cite{2021deffnerThermo,2021eisert}, quantum information \cite{2020deffnerInformation} and quantum metrology \cite{2014metrology}. The long spin lifetimes of the Dirac materials such as graphene and topological insulators make them promising candidates for quantum memory systems \cite{Chen2021}. In today's scenario, the role of non-uniform magnetic field is not limited only to the realm of physics but also has a broad perspective. For example, the non-uniformity of Earth’s magnetic field helps the migratory birds to navigate using quantum physics \cite{birds}.

We believe that the present work would motivate the interplay of quantum information theoretic ideas with the domain of relativistic quantum physics. Although, it shows the attainment of SQSL in relativistic regime, we have also included the change of QSL in variable magnetic field in presence of weak fields, which involves non-relativistic treatment. This would help in building quantum systems with low magnetic field that would have increased QSL and, thereby, could help in achieving faster quantum information processing.
The experimental probing of these facets is expected to provide rich insight into the underlying dynamics. In particular, this would have relevance to the nascent field of quantum technology.

\section*{Acknowledgment}
The authors would like to thank the referees for their constructive
comments and suggestions which helped to improve the presentation of the work. SA thanks Shivansh Aggarwal for help in presentation of Fig. \ref{fig:experiment}.
SA acknowledges Prime Minister's Research Fellows (PMRF) scheme for 
providing fellowship. 
SB acknowledges support from the Interdisciplinary
Cyber-Physical Systems (ICPS) programme of the Department of Science and 
Technology (DST), India, Grant No.: DST/ICPS/QuST/Theme-1/2019/6. SB also acknowledges 
support from the Interdisciplinary Program (IDRP) on Quantum Information and 
Computation (QIC) at IIT Jodhpur.

\section*{\label{Appendix}Appendix}
\subsection{Solution of Dirac equation in presence of magnetic field}

The Dirac equation for electron of mass $m_{e}$ and charge $q~(-e)$, in the 
presence of magnetic field is given by
\begin{equation}
i\hbar\frac{\partial\Psi}{\partial t} = \left[ c\boldsymbol{\alpha}\cdot\left(-i\hbar\textbf{$\nabla$}-\frac{q\textbf{A}}{c}\right) + \beta m_{e}c^2\right]\Psi,
\label{eqn1}
\end{equation}
where $\boldsymbol{\alpha}$ and $\beta$ are Dirac matrices, $\textbf{A}$ 
is the vector potential, $\hbar$ is the reduced Planck constant and 
$c$ is the speed of light. For stationary states, we can write
\begin{equation}
	\Psi = e^{-i\frac{Et}{\hbar}}\begin{bmatrix}
  \chi \\
 \phi \\
 \end{bmatrix}
\label{matrix},
 \end{equation} 
where $\Phi$ and $\chi$ are 2-component objects/spinors, and $\hbar$ is 
$h/2\pi$ with $h$ being Planck's constant. We consider the Pauli-Dirac representation in which
\begin{equation}
\alpha = \begin{bmatrix}
0 & \boldsymbol\sigma\\
\boldsymbol\sigma & 0\\
\end{bmatrix}
~,\beta = \begin{bmatrix}
I & 0\\
0 & -I\\
\end{bmatrix},
\end{equation}
where each block represents a $2\times 2$ matrix and $\boldsymbol{\sigma}$ 
represents three components of the Pauli matrices together in a vector.
Hence Eq. (\ref{eqn1}) reduces to
\begin{equation}
(E-m_{e}c^2)\chi = c\textbf{$\sigma$}\cdot\left(-i\hbar\textbf{$\nabla$}-\frac{q\textbf{A}}{c}\right)\phi,
\end{equation}
\begin{equation}
(E+m_{e}c^2)\phi = c\boldsymbol\sigma\cdot\left(-i\hbar\textbf{$\nabla$}-\frac{q\textbf{A}}{c}\right)\chi.
\end{equation}
Decoupling them for $\chi$, we get
\begin{equation}
(E^2-m_{e}^2c^4)\chi =\left[c\boldsymbol\sigma\cdot\left(-i\hbar\textbf{$\nabla$}-\frac{q\textbf{A}}{c}\right)\right]^2\chi.
\label{eq2}
\end{equation}
Defining $\boldsymbol{\pi}=-i\hbar\textbf{$\nabla$}-q\textbf{A}/c$ and using the identity $(\boldsymbol\sigma\cdot\boldsymbol\pi)(\boldsymbol\sigma\cdot\boldsymbol\pi) = \pi ^2 - q\hbar\boldsymbol\sigma\cdot \boldsymbol{B}/c$, Eq. (\ref{eq2}) reduces to
\begin{equation}
(E^2-m_{e}^2c^4)\chi = \left[c^2\left(\pi ^2 - \frac{q\hbar}{c}\boldsymbol\sigma\cdot \textbf{B}\right)\right]\chi, 
\label{eq3}
\end{equation}
such that the antiparticle wavefunction $\phi=-\chi$ \cite{1998rqm..book.....S}. We solve Eq. (\ref{eq3}) for our proposed 
power law variation of the magnetic field \cite{SciSris}, given by
\begin{equation}
\textbf{B} = B_{0}\rho^n \hat{z},
\end{equation}
in cylindrical coordinates $(\rho,\phi,z)$. 
Using a gauge freedom for the vector potential \textbf{A}, we choose
\begin{equation}
 \textbf{A} = B_{0}\frac{\rho^{n+1}}{n+2} \hat{\phi} = A\hat{\phi}.
\end{equation}
Hence,
\begin{equation}
  \pi ^2\chi=\left[\hat{p}_{\rho}^2+\left(\hat{p}_{\phi} - \frac{qA}{c}\right)^2+\hat{p}_{z}^2\right]\chi,
\label{eq4}
\end{equation} 
where $\hat{p}_{\rho,\phi,z}$ denote operators.
Noticing that $\phi$ and $z$ are ignorable coordinates, the solution of Eq. (\ref{eq4}) can be written as
\begin{equation}
 \chi = e^{i\left(m\phi+\frac{p_{z}}{\hbar}z\right)}R(\rho),
 \end{equation}
where $R(\rho)$ is a two-component matrix, `$m\hbar$' is the angular momentum 
of the electron and $p_{z}$ is the eigenvalue of momentum in the $z-$direction. Therefore, Eq. (\ref{eq4}) becomes
\begin{multline}
\pi ^2 R = -\hbar ^2 \left[\frac{\partial ^2}{\partial \rho^2}+\frac{1}{\rho}\frac{\partial}{\partial\rho}-\frac{m^2}{\rho^2}\right]R(\rho)\\ 
+\left[\frac{q^2A^2}{c^2}+\frac{2q\hbar mA}{c\rho}+p_{z}^2\right]R(\rho).
\label{eq5}   
\end{multline} 
From Eqs. (\ref{eq3}), (\ref{eq4}) and (\ref{eq5}) and substituting $q = -e$, we obtain
\begin{eqnarray}
\nonumber
&&\left(\frac{E^2 - m_{e}^2c^4}{c^2}-p_{z}^2\right)R(\rho) =  -\hbar ^2 \left[\frac{\partial ^2}{\partial \rho^2}+\frac{1}{\rho}\frac{\partial}{\partial\rho}-\frac{m^2}{\rho^2}\right]R(\rho)\\
&&\qquad+\left[\frac{e^2A^2}{c^2}-\frac{2e\hbar mA}{c\rho}+\frac{e\hbar}{c}(\sigma_{z}B)\right]R(\rho).
\label{eq6}
\end{eqnarray}
There will be two independent solutions for $R(\rho)$, which can be taken, without loss of generality, to be the eigenstates of $\sigma_{z}$, with eigenvalues $\pm 1$. Thus if we choose two independent solutions of the form
\begin{equation}
 R_{+}(\rho)=\begin{bmatrix}
 \tilde{R}_{+}(\rho)\\
 0
 \end{bmatrix}
~,R_{-}(\rho)=\begin{bmatrix}
 0\\
 \tilde{R}_{-}(\rho)\\
 \end{bmatrix}
 \nonumber
 \end{equation}
such that $\sigma_{z}{R}_{\pm}=\pm{R}_{\pm}$, Eq. (\ref{eq6}) becomes

\begin{multline}
\tilde{P}\tilde{R}_{\pm}= -\hbar ^2 \left[\frac{\partial ^2}{\partial \rho^2}+\frac{1}{\rho}\frac{\partial}{\partial\rho}-\frac{m^2}{\rho^2}\right]\tilde{R}_{\pm}\\+\left[\frac{e^2A^2}{c^2}-\frac{2e\hbar mA}{c\rho}\pm\frac{e\hbar}{c}B\right]\tilde{R}_{\pm} \label{eq8}
\end{multline}
where
\begin{equation}
\tilde{P} = \left(\frac{E^2 - m_{e}^2c^4}{c^2}-p_{z}^2\right).
\label{eqP}
\end{equation}
Dividing Eq. (\ref{eq8}) by $m_{e}^2c^2$, we have an eigenvalue equation as
\begin{align}
\alpha\tilde{R}_{\pm} &= -\left(\frac{\hbar}{m_{e}c}\right)^2 \left[\frac{\partial ^2}{\partial \rho^2}+\frac{1}{\rho}\frac{\partial}{\partial\rho}-\frac{m^2}{\rho^2}\right]\tilde{R}_{\pm}  \nonumber\\
&\quad+\left[\frac{e^2A^2}{m_{e}^2c^4}+\frac{e\hbar}{m_{e}^2c^3}\left(-\frac{2mA}{\rho}\pm B\right)\right]\tilde{R}_{\pm}\nonumber \\ 
  &= -\lambda_{e}^2\left[\frac{\partial ^2}{\partial \rho^2}+\frac{1}{\rho}\frac{\partial}{\partial\rho}-\frac{m^2}{\rho^2}\right]\tilde{R}_{\pm}
\nonumber \\
&\quad  + \left[\left(\frac{kB_0\rho^{n+1}}{n+2}\right)^2 + k\lambda_{e}\left(-\frac{2m}{n+2}\pm 1\right)B_0\rho^n\right]\tilde{R}_{\pm},
\label{eq10}
\end{align}
where 
$\epsilon = \frac{E}{m_{e}c^2}$ (dimensionless energy),
$x_{z} = \frac{p_{z}}{m_{e}c}$\,\; (dimensionless momentum along $z-$direction),
$\lambda_{e} = \frac{\hbar}{m_{e}c}$ (Compton wavelength of electrons),
$k=\frac{e}{m_{e}c^2}$ and $\alpha = \frac{\tilde{P}}{m_{e}^2 c^2} = (\epsilon^2 - 1-x_{z}^2)$,
is the square of dimensionless energy and acts as an eigenvalue of the problem \cite{SciSris}.
Therefore, energy of level $\nu$ becomes
\begin{equation}
    E_{\nu} = m_e c^2 \sqrt{1+\alpha_{\nu}}.
    \label{EalphaEq}
\end{equation}

For uniform magnetic field, i.e. $n=0$, from the solution of Eq. (\ref{eq10}), the eigenvalue of level $\nu$ is given by
\begin{equation}
\alpha_{\nu} = 2k\lambda_{e}B_{0}\left (\nu+\frac{|m|}{2}-\frac{m}{2}+\frac{1}{2}\pm\frac{1}{2}\right).
\label{eq12}
\end{equation}
Hence, in the limiting case of $B_0\rightarrow\infty$ for $m=0$, except for ground state energy of spin-down electron which is the rest mass energy $m_ec^2$, energy of level $\nu$ is given by
\begin{eqnarray}
\Lim{B_0\rightarrow\infty}E_{\nu}&&\rightarrow m_e c^2\sqrt{\alpha_{\nu}}\\
&& = m_e c^2 2 \lambda_e \beta \sqrt{\left (\nu+\frac{1}{2}\pm\frac{1}{2}\right)},
\label{lim energy uniform}
\end{eqnarray}
where
\begin{equation}
    \beta = \sqrt{\frac{kB_0}{2\lambda_e}}.
\end{equation}

\subsection{\label{Apppendix2}Estimation of Magnetic field in the proposed experimental design}

For an electromagnet with airgap between pole pieces, magnetic field is given by
\begin{equation}
    B = \frac{NI\mu_0 \mu}{L_c \mu_0 + L_G \mu} \approx \frac{K}{L_G}.
    \label{eq:bAir}
\end{equation}
where $N$ is the number of turns, $I$ is current in the wire, $L_c$ is the magnetic length of the core, $L_G$ is the gap between the pole pieces, $\mu_0$ is vacuum permeability, $\mu~(>>\mu_0)$ is the permeability of core and $K$ is a constant. 

 
Let us assume that the surface of the pole pieces is determined by the equation
\begin{equation}
    z = z_0(r+r_0)^{\beta},
    \label{eq:surface}
\end{equation}
such that $z_0$ is the scaling factor and $r_0$ is a length scale, which is very small compared to the size the circular plane confining the electron, determining the variation of the surface close to the origin. Such a surface can be easily engineered with appropriate machining of the soft iron core.
Noting $L_G$ is nothing but twice the $z$-coordinate, and combining Eqs. (\ref{eq:bAir}) and (\ref{eq:surface}) (neglecting edge effects), we get 
\begin{equation}
    B(r) = \frac{K}{2z_0(r+r_0)^{\beta}} \approx B_0 r^n
    \label{eq:Bexp}
\end{equation}
with $n=-\beta$.
Thus, the above equation shows how one can achieve non-uniform magnetic field using the proposed experimental design in \S\ref{exp}.
\bibliographystyle{apsrev4-1}
\bibliography{quantum}
\end{document}